\documentstyle[11pt,amssymb,amsfonts,amsthm,amssymb,amscd,amstext,epsfig]{article}  

\newcommand{\startappendix}{
\setcounter{section}{0}
\renewcommand{\thesection}{\Alph{section}}}
\newcommand{\Appendix}[1]{
\refstepcounter{section}
\begin{flushleft}
{\large\bf Appendix \thesection: #1}
\end{flushleft}}
\def\ben{\begin{equation}}
\def\een{\end{equation}}

\let\a=\alpha    
  \let\q=\theta 
     
\let\s=\sigma

\def\be{\begin{equation}}
\def\ee{\end{equation}}
\def\ba{\begin{array}}
\def\ea{\end{array}}

\def\dalemb#1#2{{\vbox{\hrule height .#2pt
        \hbox{\vrule width.#2pt height#1pt \kern#1pt
                \vrule width.#2pt}
        \hrule height.#2pt}}}

\newcommand{\bea}{\begin{eqnarray}}
\newcommand{\eea}{\end{eqnarray}}

\newcommand{\Tr}{{\rm Tr} }

\def\R{{{\Bbb R}}}

\def\N{{{\Bbb N}}}

\begin{document}
\begin{flushright}
\hfill{DAMTP-2005-30} \\
{hep-th/0503238}
\end{flushright}

\begin{center}
\vspace{1cm}
{ \LARGE {\bf The $O(N)$ model on a squashed $S^3$ and the
    Klebanov-Polyakov correspondence}}

\vspace{1.5cm}

Sean A. Hartnoll${}^1$ and S. Prem Kumar${}^{1,2}$

\vspace{0.8cm}

{\it ${}^1$ DAMTP, Centre for Mathematical Sciences,
Cambridge University\\ Wilberforce Road, Cambridge CB3 OWA, UK}

\vspace{0.3cm}

{\it ${}^2$ Department of Physics, University of Wales Swansea\\
Swansea, SA2 8PP, UK}

\vspace{0.3cm}

s.a.hartnoll@damtp.cam.ac.uk \hspace{1cm} p.kumar@damtp.cam.ac.uk

\vspace{2cm}

\end{center}

\begin{abstract}

We solve the $O(N)$ vector model at large
$N$ on a squashed three-sphere with a conformal mass term. Using the
Klebanov-Polyakov version of the AdS$_4$/CFT$_3$ correspondence we
match various aspects of the strongly coupled theory with the physics
of the bulk AdS Taub-NUT and AdS Taub-Bolt geometries. Remarkably, we
find that the field theory reproduces the behaviour of the bulk free
energy as a function of the squashing parameter. The $O(N)$ model is realised in a
symmetric phase for all finite values of the coupling and squashing
parameter, including when the boundary scalar curvature is negative.  

\end{abstract}

\pagebreak
\setcounter{page}{1}
\section{Introduction and summary}

The AdS/CFT correspondence posits the existence of exact dualities between large
$N$ field theories and string theory on asymptotically Anti-de
Sitter (AdS) spacetimes
\cite{Maldacena:1997re,Witten:1998qj,Gubser:1998bc}. As is well
known, these gauge-gravity dualities typically relate two complex
theories, each being tractable only in some limiting region of
parameter space. A somewhat simpler version of such a duality was
recently proposed by Klebanov and Polyakov \cite{Klebanov:2002ja}
wherein the large $N$ field theory is exactly solvable.
Specifically, the Klebanov-Polyakov correspondence conjectures a
duality between a theory of massless higher spin gauge fields in
${\rm AdS}_4$ spacetime on the one hand, and the singlet sector of
the critical $O(N)$ vector model at large $N$ in three dimensions
on the other.

Motivated by the Klebanov-Polyakov proposal, in this paper we will
solve the $O(N)$ vector model at large $N$ on homogeneous,
compact manifolds known as squashed three-spheres. The squashed
three-spheres are one parameter deformations of the round
three-sphere which introduce an anisotropy whilst preserving
homogeneity. Importantly, they are the conformal boundary of
certain asymptotically locally Anti-de Sitter spacetimes known as
AdS Taub-NUT and, for certain values of the squashing parameter,
of another geometry called AdS Taub-Bolt
\cite{Chamblin:1998pz,Hawking:1998ct}. Both these backgrounds are
Einstein manifolds with constant negative curvature. The
holographic correspondence then suggests an equivalence between
higher spin gauge theory on the AdS Taub-NUT/Bolt geometries, and
the $O(N)$ vector model at large $N$ on the boundary geometry.

These backgrounds were studied in the context of the ${\rm
AdS}_4/{\rm CFT}_3$ correspondence in
\cite{Chamblin:1998pz,Hawking:1998ct,Emparan:1999pm,Mann:1999pc,Zoubos:2002cw,
Zoubos:2004qm,Kleban:2004bv,Astefanesei:2004kn}
. However, in those works little could be said about the dual
field theory because in the $AdS_4\times S^7$ version of the
gauge-gravity correspondence \cite{Maldacena:1997re} the field
theory is a strongly coupled conformal field theory which remains
largely unknown. This is no longer the case for the
Klebanov-Polyakov version of the correspondence, where the field
theory is tractable at all values of the coupling, weak and strong.

We will study the $O(N)$ theory on a squashed three-sphere as a
function of the squashing parameter and a dimensionless coupling
constant. In the absence of a complete formulation of the theory
of higher spin gauge fields on the bulk geometries of interest, we
compare exact results from field theory with semiclassical
properties of Einstein gravity in the bulk
\cite{Chamblin:1998pz,Hawking:1998ct,Emparan:1999pm,Mann:1999pc,Kleban:2004bv}.
One of the main results of our study is a remarkable qualitative
agreement between the free energy of the field theory at strong
coupling and that of the bulk gravitational theory, as a function
of the squashing parameter. This result is encapsulated in figures 3
and 4 below.

The squashed sphere is an $S^1$ bundle over $S^2$
where the squashing parameter $\alpha$ is related to the
periodicity of the $S^1$ fibre. The periodicity of the $S^1$ fibre
is often interpreted as the inverse of a temperature, giving
$T\sim\sqrt{\alpha}$ for large $\alpha$. This then allows a
physical understanding of the fact that the bulk gravity theory
with a squashed three-sphere as boundary exhibits a first order
phase transition of Hawking-Page type \cite{Hawking:1982dh} as
this temperature is increased beyond a critical value
\cite{Chamblin:1998pz,Hawking:1998ct}. In particular, for fixed
$\alpha <\alpha_{\rm crit}$ the semiclassical quantum gravity
partition function is dominated by the AdS Taub-NUT geometry.
Beyond $\alpha_{\rm crit}$ however, the partition function is
dominated by the AdS Taub-Bolt background. The AdS Taub-Bolt
geometry should be thought of as a Euclidean black hole with a NUT
charge.

Our field theory result for the free energy of the $O(N)$ vector
model at strong coupling will agree with the behaviour of AdS
Taub-NUT for $\alpha< \alpha_{\rm crit}$ and with that of AdS
Taub-Bolt at large values of $\alpha$, but with a smooth crossover
between the two regimes instead of a first order phase transition. We view
the qualitative agreement away from $\alpha=\alpha_{\rm crit}$ as
a positive test of the proposed holographic duality. We will
discuss the possibility that accounting for the infinite massless
higher spin degrees of freedom in the bulk will smooth out the
first order Hawking-Page transition.

It is not surprising that the nature of the Hawking-Page
transition here is qualitatively different to the cases in which
it is dual to confinement/deconfinement transitions in large $N$
Yang-Mills theory
\cite{Witten:1998qj,Aharony:2005bq,Aharony:2003sx,Alvarez-Gaume:2005fv}.
These transitions are accompanied by a drastic change in the
number of degrees of freedom of the boundary field theory. The
natural phase transition one might expect in the $O(N)$ model is a
symmetry breaking transition as one varies $\alpha$. However, we
will show explicitly that the large $N$ limit of the $O(N)$ model
on a compact space is always realized in an $O(N)$-symmetric phase
at finite coupling, with any possibility of spontaneous symmetry
breaking of $O(N)\rightarrow O(N-1)$ precluded by finite volume
effects.

Let us now briefly summarise a few further features and
implications thereof of our field theoretic results. In flat
space, the critical $O(N)$ vector model in three dimensions
exhibits a renormalisation group flow from a free fixed point in
the ultraviolet (UV) to an interacting fixed point in the infrared
(IR), see for example
\cite{Klebanov:2002ja,Moshe:2003xn,zinn-justin}. This flow is
induced by the quartic interaction which is a relevant operator
with a coupling constant that has mass dimension one. A compact
space such as the squashed sphere introduces a second scale, thus
providing a dimensionless coupling constant. The large $N$ limit
leads to an exactly solvable theory for any value of this
dimensionless coupling constant.

We consider the theory with a conformal mass term so that in the
ultraviolet the theory approaches a free conformal fixed point,
presumably corresponding to the asymptotically AdS region of the
bulk. Interestingly, the conformal mass term becomes negative for
a range of values of the squashing, as the scalar curvature of the
squashed sphere becomes negative. One of our results will be that
nonetheless the theory is always realised in a symmetric phase for
all finite values of the coupling and squashing.

In addition to matching the behaviour of the free energy with
squashing as we have described above, we make two further
connections between the physics of the $O(N)$ model and the bulk theory
\begin{itemize}

\item A negative curvature of the boundary is often associated with
  instabilities in the bulk geometry, as reviewed in
  \cite{Kleban:2004bv}. Our field theory results suggest that there
  is no instability of the classical bulk AdS Taub-NUT geometry in the
  higher spin gauge theory when the scalar curvature of the conformal
  boundary becomes negative. We will perform a check of this statement
  by considering fluctuations of a conformally coupled scalar field
  about the bulk background.

\item We calculate the condensate
  $\langle\vec{\Phi}{\cdot}\vec{\Phi}\rangle$ in
  the $O(N)$ model. This vacuum expectation value is dual, in the
  Klebanov-Polyakov correspondence, to the normalisable mode of a
  conformally coupled scalar field in the bulk, $\varphi$. This scalar
  field is not turned on in the AdS Taub-NUT or AdS Taub-Bolt
  geometries. Consistent with this fact we find
  $\langle\vec{\Phi}{\cdot}\vec{\Phi}\rangle \to 0$ at strong coupling. For
  boundaries other than the round sphere, this condensate does not vanish at
  finite coupling, predicting curvature corrections to the bulk AdS
  Taub-NUT/Bolt geometries, whilst AdS itself is protected from corrections.
\end{itemize}

Given that the nonperturbative large $N$ resummation can be
implemented in field theory for arbitrary values of the
dimensionless coupling, we can in fact analyse the theory both at
arbitrarily weak coupling and at infinitely strong coupling. In
these two limits we find analytically the following field theory
physics
\begin{itemize}
\item The theory at arbitrarily weak dimensionless coupling,
  $a\lambda \to 0^+$, exhibits a phase
  transition at a value of the squashing parameter where the boundary
curvature
  turns negative. The order parameter for this transition is the
  condensate $\lambda \langle\vec{\Phi}{\cdot}\vec{\Phi}\rangle$ which 
vanishes for
negative curvature
  while remaining nonzero and of order $N$ at positive
  curvatures. Interestingly, for any finite value of the coupling this
  phase transition gets completely smoothed out, as we stated above.
  Such nonperturbative effects may be relevant more generally for
  attempts to extrapolate perturbative phase transitions to strong
  coupling \cite{Aharony:2005bq,Aharony:2003sx}.

\item At large squashing parameter the free energy of the AdS
  Taub-Bolt geometries is known to depend linearly on the squashing
  parameter \cite{Chamblin:1998pz,Hawking:1998ct}. We find that the
  $O(N)$ model exhibits the same linear dependence at both strong and
  weak coupling. The ratio of the free energy at large squashing at weak and 
strong
  coupling is 4/5, rather analogously to the well-known 3/4 factor
  \cite{Gubser:1996de} for
  ${\mathcal{N}}=4$ super Yang Mills, except that here we can calculate
  both limits in field theory.
\end{itemize}

The physics of interacting quantum fields on a squashed
three-sphere received attention at various points in the eighties
following the growth of interest in quantum fields on curved
backgrounds. These works included considerations of quantum
induced symmetry breaking through calculation of the one loop
effective potential \cite{Critchley:1981am,Critchley:1982ch,
  Shen:1985ir,Okada:1985es,Shen:1986jr}.
However, although one loop effective potentials can be useful,
they are insufficient to understand the phase structure of field
theories. Firstly because when quantum corrections become of the
same order as the classical mass term then typically perturbation
theory is not reliable. Secondly, the one loop minima of the
effective potential are generically located at a nonperturbative
mass scale, again invalidating the reliability of the perturbative
analysis. Our results described above for negative values of the
conformal mass term highlight the inadequacy of a perturbative
treatment.

The layout of this paper is as follows. Section \ref{sec:model}
introduces the $O(N)$ model on a general background, including
effects due to compactness of the spacetime. Section
\ref{sec:sis1} specialises to the squashed three-sphere and
studies the conformally coupled theory. Section \ref{sec:dual}
connects the field theory results with the dual gravitational
description using the Klebanov-Polyakov correspondence. The
appendices contain technical calculations of the zeta function on
a squashed three-sphere, as well as a check on our calculations.

\section{The $O(N)$ model at large $N$}
\label{sec:model}

We begin by reviewing known results on the interacting $O(N)$
scalar field theory on a $D$ dimensional spacetime ${\mathcal{M}}$
(see {\it e.g.} \cite{Moshe:2003xn}). We also note some subtleties
arising from the fact that the spacetime ${\mathcal{M}}$ is compact.
The large $N$ limit of the
field theory will allow us to analyse the exact effective
action for this theory and thus draw reliable conclusions about
the dynamics of the quantum theory. In Euclidean signature the
$O(N)$ model in $D$ dimensions has the classical action
\be\label{eq:scl}
{S_{cl}[\vec\Phi]=\int d^D x \sqrt{g}\left[ {1\over
      2}{\nabla}\vec{\Phi}{\cdot}{\nabla}\vec{\Phi} + {1\over 2}
\,m^2\, \vec \Phi^2 + \frac{\lambda}{4 N} (\vec\Phi\cdot\vec\Phi)^2 \right]}
\,,
\ee
where $\vec \Phi$ is an $N$-component field transforming as a
vector under $O(N)$ rotations. The interacting scalar field theory
is super-renormalisable in dimensions less than four.
The explicit
$N$ dependence in (\ref{eq:scl}) is necessary for the theory to
posses a well defined large $N$ limit. In the action
(\ref{eq:scl}) the mass term includes a possible $\xi R$ coupling
to the scalar curvature of the background.

To solve the large $N$ theory one introduces an auxiliary scalar
field $\sigma$ which permits a rewriting of the action as
\be\label{eq:aux1}
{S_{cl}[\vec\Phi]=\int d^D x \sqrt{g}\left[ {1\over
      2}{\nabla}\vec{\Phi}{\cdot}{\nabla}\vec{\Phi} + \frac{1}{2}
      \left(m^2 + \lambda \sigma \right) \vec\Phi\cdot\vec\Phi -
      \frac{\lambda N}{4} \sigma^2 \right]} \,.
\ee
The field $\sigma$ is the composite operator,
\be\label{eq:aux2}
{N \sigma=\vec\Phi\cdot\vec\Phi\,.}
\ee

To derive the large $N$ effective potential we introduce a homogeneous
background
expectation value for the $O(N)$ field and integrate out all inhomogeneous
fluctuations about this configuration. Without loss of generality we
may always choose to rotate this expectation value into the top
component of the $O(N)$ vector, so that
\be\label{eq:bf}
{\vec\Phi = ({\sqrt N}\phi +\delta\phi,\;\pi_1, \;\pi_2,\ldots,
  \;\pi_{N-1})} \,,
\ee
where $\phi$ is a homogeneous background and $\delta\phi$ and $\vec\pi$
are the quantum fluctuations.
As we will see shortly, the explicit factor
of $\sqrt N$ in (\ref{eq:bf}) is the correct scaling behaviour for VEVs in
the
large $N$ theory so that $\phi \sim {\cal O}(N^0)$.
At this point it is worth noting that the backgrounds we consider in
this paper, squashed three-spheres, are homogeneous and hence
the notion of an effective potential for homogeneous fields
makes sense.

Introducing a homogeneous background might appear to break the $O(N)$
symmetry to {$O(N-1)$}. However, the path integral must include an integral 
over the
vacuum manifold and we will see that this implies that
symmetry breaking does not actually occur on a
compact space. In fixing the $O(N)$ symmetry for the
homogeneous modes in (\ref{eq:bf}) we have effectively transformed into
polar coordinates for $\vec\Phi$ with $\sqrt N \phi$ playing the role
of the radial coordinate. Upon integrating over the angular directions
in field space the partition function picks up
an extra factor given by the volume of the vacuum manifold
$O(N)/O(N-1) = S^{N-1}$
\be\label{eq:volume}
\frac{2 \pi^{(N-1)/2}}{\Gamma[(N-1)/2]}\left(N^{1/2} \phi \right)^{N-1}
\; \stackrel{N \to \infty}{=} \;
\phi^N \pi^{N/2} e^{N/2} = \exp\left({\displaystyle N/2\left[1 + \ln\pi + 
\ln
    (\phi^2/\mu) \right]}\right)\,,
\ee
where $\mu$ is an arbitrary scale introduced to keep the arguments of
the logarithms dimensionless. We note that there is a curious cancellation 
between
the $N^N$ term coming from the radius of the $N$-sphere and from the large
$N$ limit of the Gamma function. The result of this cancellation is that the
extra contribution to the partition function (\ref{eq:volume}) has the
same $N$ dependence as the remaining terms in the effective action in
the large $N$ limit.

In the large $N$ limit the effective potential for the
homogeneous fields $\phi$ and $\sigma$, after integrating out
$\vec\pi$ and including the contribution (\ref{eq:volume}),
is \footnote{The contribution from integrating out
  the fluctuations $\delta\phi$ is subleading in the $1/N$ expansion.}
\bea\label{eq:veff}
\frac{V_\text{eff}(\phi, \sigma)}{N} & = & {1\over 2} \left(m^2+\lambda
\sigma\right) \phi^2 - \frac{\lambda}{4} \sigma^2 + {1\over{2\rm Vol {\cal
M}}}\ln
\det{}'\left[{-\Box +m^2 + \lambda \sigma \over\mu^2}\right]
\nonumber \\
& - & \frac{1}{2 \rm Vol {\cal M}} \left[1 + \ln \pi + \ln
  \frac{\phi^2}{\mu} \right] \,,
\eea
where ${\rm Vol{\cal M}}$ is the volume of the spacetime and $\mu$
is again an arbitrary dimensionful scale which should be interpreted as
a sliding renormalisation scale. The prime in $\det{}'$ denotes
the fact that the integration over the $\vec\pi$ fluctuations did not
include the constant modes, which we have dealt with separately.

In the large $N$ limit only the saddle point configuration
obtained by extremising (\ref{eq:veff}) contributes to the
partition function. Extremising $V_\text{eff}(\phi,\sigma)$ with
respect to the background fields $\phi$ and $\sigma$ we then find
the vacuum equations
\be\label{eq:sp0}
\phi^2 (m^2 + \lambda \sigma) = \frac{1}{\rm Vol{\cal M}} \,,
\ee
and
\be\label{eq:sp1}
{\left({\phi^2}- \sigma \right) + {1\over {\rm Vol}{\cal M}} \Tr{}'
\left[\frac{1}{-\Box + m^2 + \lambda \sigma}\right]  =0\,.}
\ee
The second of these equations
is in fact, in the large $N$ limit, just the vacuum expectation value of
the operator equation (\ref{eq:aux2}) which would yield $\sigma =
\phi^2 + \langle\vec\pi^2\rangle/N$. Here $\vec\pi$ refers to the
$(N-1)$ ``pions'' of equation (\ref{eq:bf}). Thus the functional
trace computes the renormalised vacuum expectation value of the
composite operator $\vec\pi\cdot\vec\pi$, encoding the
contribution of the quantum fluctuations about the ground state.

At this stage it is convenient to define the ``effective pion
mass''
\be\label{eq:mass}
{m_\pi^2 = m^2 + \lambda \sigma \,,}
\ee
which is precisely the mass of the $(N-1)$ ``pion'' fluctuations
in (\ref{eq:bf}). From equation
(\ref{eq:sp1}) we see that this effective mass incorporates a
self-consistent resummation of the quantum fluctuations about the
vacuum state. This is a consequence of the large $N$ limit which
resums the so-called cactus diagrams of the theory.

The effective mass $m_\pi^2$ depends on whether the theory is
realised in a symmetric phase or not. In an $O(N)$-symmetric
phase $m_\pi^2$ is nonvanishing. In a symmetry
broken phase we would expect $m_\pi^2$
to vanish, thus implementing Goldstone's theorem. That is, in a
symmetry broken phase there would be $N-1$ massless Goldstone bosons.
In equation (\ref{eq:sp0}) we see that $m_\pi^2$ cannot be zero.
To achieve this we would need to take ${\rm Vol{\cal M}}$ or $\phi^2$ to
infinity. We will see below that taking $\phi^2$ to infinity requires
taking the coupling to zero. Thus we find that
symmetry breaking cannot occur for the $O(N)$ model
in a compact spacetime at large $N$ and finite coupling.

Equations (\ref{eq:sp0}) and (\ref{eq:sp1}) may be rewritten as a gap
equation for $m_\pi^2$
\be\label{eq:pionmass}
\fbox{ $\displaystyle
m_\pi^2 = m^2 + \frac{\lambda}{\rm Vol{\cal M}} \Tr
\left[\frac{1}{-\Box + m_\pi^2}\right] \,. $}
\ee
Note that the trace now includes the constant mode, as will all
remaining traces in this paper. Once we have solved
(\ref{eq:pionmass}) for $m_\pi^2$ we may evaluate the effective
potential at the extremum to find the action
\bea\label{eq:F}
I & = & \int d^Dx \sqrt{g}\, V_{\text{eff.}} \nonumber \\
& = & \frac{N}{2} \left[ -
  \frac{{\rm Vol{\cal M}}}{2 \lambda} (m^2-m_\pi^2)^2 + \ln
\det\left(\frac{-\Box +
    m_\pi^2}{\mu^2} \right) + \ln (\mu^3 {\rm Vol{\cal M}}) \right] +
\text{const.}
\eea
The only effect of scaling $\mu$ is to change the unphysical additive
constant in the action.

\subsection{Dynamics on a curved background ${\mathcal{M}}$}

To compute the large $N$ effective potential and its minima, we
need to evaluate the functional traces of equations
(\ref{eq:pionmass}) and (\ref{eq:F})
on ${\mathcal{M}}$. A natural way of doing this is by using zeta
function regularisation. Recall that the zeta function of an elliptic
operator $A$ is defined by
\be\label{eq:defzeta}
\zeta(s) = \Tr A^{-s} \,.
\ee
Thus for the operator $[-\Box +
m_\pi^2]/\mu^2$ on ${\mathcal{M}}$ we set
\be
{ \ln \det\left(\frac{-\Box + m_\pi^2}{\mu^2} \right)= -\lim_{s\rightarrow
0} {d\over
ds}\Tr \left(\frac{-\Box + m_\pi^2}{\mu^2} \right)^{-s} \equiv -
\zeta^\prime(0)} \,.
\ee
Differentiating this expression one obtains
\be\label{eq:tracezeta}
\Tr \left[\frac{1}{-\Box + m_\pi^2}\right]
= \frac{1}{\mu^2} \lim_{s\rightarrow 0}{d\over ds} (s\;\zeta(s+1)) \,.
\ee

The main technical aspects of this work will involve the computation
of the zeta function on a squashed three-sphere. The methods we use
are similar to those in \cite{Shen:1986jr}, although we are working in
a nonperturbative framework.

\section{$O(N)$ model on a squashed three-sphere}
\label{sec:sis1}

In this section we apply the above formalism with a squashed
three-sphere as the background.
The metric of the squashed three-sphere of radius $a$ is
\be\label{eq:metric}
ds^2 = \frac{a^2}{4} \left(\s_1^2 + \s_2^2 + \frac{1}{1+\a}
  \s_3^2 \right) \,,
\ee
where $\s_i$ are the usual left-invariant $SU(2)$ one forms
\be
\s_1 + i \s_2 = e^{-i\psi} \left(d\q+i\sin\q d\phi \right) \,,\quad
\s_3 = d\psi + \cos\q d\phi \,,
\ee
and $\a$ is the squashing parameter
which may take values in the range $\a \in (-1,\infty)$.
In this parameterisation $\a=0$ is the round three-sphere.
The other manifolds are also topologically spheres, but with a
squashed $S^1$ fibration over $S^2$. The scalar
curvature is found to be
\be\label{eq:Ricci}
R = \frac{2(3+4\a)}{a^2(1+\a)} \,.
\ee
Note that the scalar curvature changes sign and becomes negative
when $\a < -3/4$. We will also need the volume
\be
\text{Vol}({\mathcal{M}}) = \frac{2 a^3 \pi^2}{(1+\a)^{1/2}} \,.
\ee

Although squashing the round sphere introduces an anisotropy, it
preserves the homogeneity of the space.
Hence the spectrum of the scalar Laplacian on (\ref{eq:metric}) is
straightforward to calculate. As we saw above, this is what we need in
order to compute the quantum correction to the self-energy which then
enters the right hand side of the gap equation (\ref{eq:pionmass}) of
the large $N$ theory on the squashed sphere.
Following some minor manipulations
\cite{Shen:1986jr}, the zeta function (\ref{eq:defzeta}) for the
operator $(-\Box+m^2_\pi)/\mu^2$ on the squashed sphere may be written as
\be\label{eq:sum}
\zeta(s) = \sum_{l=1}^{\infty} \sum_{q=0}^{l-1} \frac{l
(a\mu)^{2s}}{\left[l^2+\a
  (l-1-2q)^2 + a^2m^2_\pi - 1\right]^s} \,.
\ee
We may obtain a finite expression for $\zeta(s)$ by
analytic continuation of the sum (\ref{eq:sum}).
The methods are
standard and are described in Appendix A. We find
\bea\label{eq:zetas}
\frac{1}{\mu^{2s}}\zeta(s) & = & \frac{1}{m^{2s}_\pi} +
\frac{4a^{2s}}{(3+a^2 m^2_\pi+\a)^s} + a^{2s} \Theta(\a) H \nonumber \\
& + & a^{2s} \int_0^1 \frac{F(y) dy}{(1+\a y^2)^s} +
\frac{a^{2s}}{(1+\a)^s} \int_0^{\infty} \frac{G(y) dy}{e^{2\pi y} + 1} \,,
\eea
where the functions $F,G$ and $H$ are given in Appendix A and
$\Theta(\a)$ is the Heaviside step function. We point out that
although the appearance of the step function might suggest a
discontinuity in the zeta function at $\alpha=0$, the function $H$ and
all its derivatives actually vanish exponentially at $\alpha=0$, see equation
(\ref{eq:defh}) below. In fact the connection between the squashed
sphere and thermal field theory is seen explicitly in the
function $H$ and $dH(s)/ds|_{s=0}$ (see Appendix A). The latter has a
natural interpretation as the finite temperature free energy of a
field theory on $S^2$ accompanied by certain parity reversals.

For technical reasons in performing the analytic continuation, again see
Appendix A, we have restricted ourselves to $-8/9 < \a < \infty$,
omitting the range $-1 < \a \leq -8/9$. The range we consider includes the
value
$\a=-3/4$, where the Ricci scalar changes sign, the value
corresponding to the round sphere, $\a=0$, and also
arbitrarily large values of $\a$ that are relevant for the Hawking-Page
type transition in the dual gravitational theory. From our results, the
extrapolation to the range $-1\leq\alpha<-8/9$ seems straightforward,
although the $\a \to -1$
limit itself is
fairly singular and may contain interesting physics that deserves to be
studied separately.

\subsection{Physics of the symmetric vacuum I: the mass gap}

In three dimensions, the coupling constant $\lambda$ has mass
dimension one and thus the interaction is a relevant operator which
generates a nontrivial renormalisation group flow from a free fixed
point in the UV. High energies correspond to weak coupling, and at
long wavelengths the coupling becomes stronger. On the squashed
sphere with scale size $a$ a natural dimensionless coupling
$a\lambda$ can be defined. We investigate the theory at different
values of this dimensionless coupling, including arbitrarily strong
and weak coupling regimes. Varying this coupling naturally corresponds
to varying the size of the compact space keeping $\lambda$ fixed. For
any value of the coupling, the IR physics of the interacting large $N$
theory is encapsulated in the solution to the mass gap equation
(\ref{eq:pionmass}).

In three dimensions, zeta functions are regular at $s=1$.
Therefore the right hand side of
expression (\ref{eq:tracezeta}) reduces to $\zeta(1)$. The gap
equation (\ref{eq:pionmass}) then becomes
\be\label{eq:massivesol}
m^2_\pi = m^2 + \frac{\lambda}{\mu^2 \text{Vol}{\mathcal{M}}}
\zeta_{m_\pi^2}(1) \,.
\ee

To ensure that in the ultraviolet our theory flows to a free
conformal fixed point, we set the bare mass term, $m^2$, to be given
by the conformal coupling to the background curvature $R$
\be
m^2 = m^2_{\text{conf.}} = \frac{1}{8} R = \frac{3+4\alpha}{4 a^2
(1+\alpha)}\,.
\ee
The gap equation may now be written as
\be
a^2 m^2_\pi = \frac{3+4\a}{4(1+\a)} + \frac{a\lambda
  (1+\a)^{1/2}}{2\pi^2} \frac{\zeta_{m_\pi^2}(1)}{(a\mu)^2} \,.
\ee
The only dependence on $a$, the size of the squashed sphere, in
this expression is contained in the
dimensionless masses and couplings $a\lambda$ and $a^2 m_\pi^2$.

It is clear that (\ref{eq:massivesol}) always has
solutions with $m_\pi^2 > 0$, essentially due to the $1/m_\pi^2$ behaviour 
in
$\zeta_{m_\pi^2}(1)$ as $m_\pi^2\to 0$. In figure 1 we plot the solution for
$a^2 m_\pi^2$ as a
function of $\a$ for three fixed values of $a\lambda$. The
existence of a symmetric vacuum for all couplings $a\lambda$ could
only be seen because we have solved the theory nonperturbatively.
At a perturbative level there is no symmetric phase for $\a <
-3/4$ and small $a\lambda$.
\begin{figure}[h]
\begin{center}
\epsfig{file=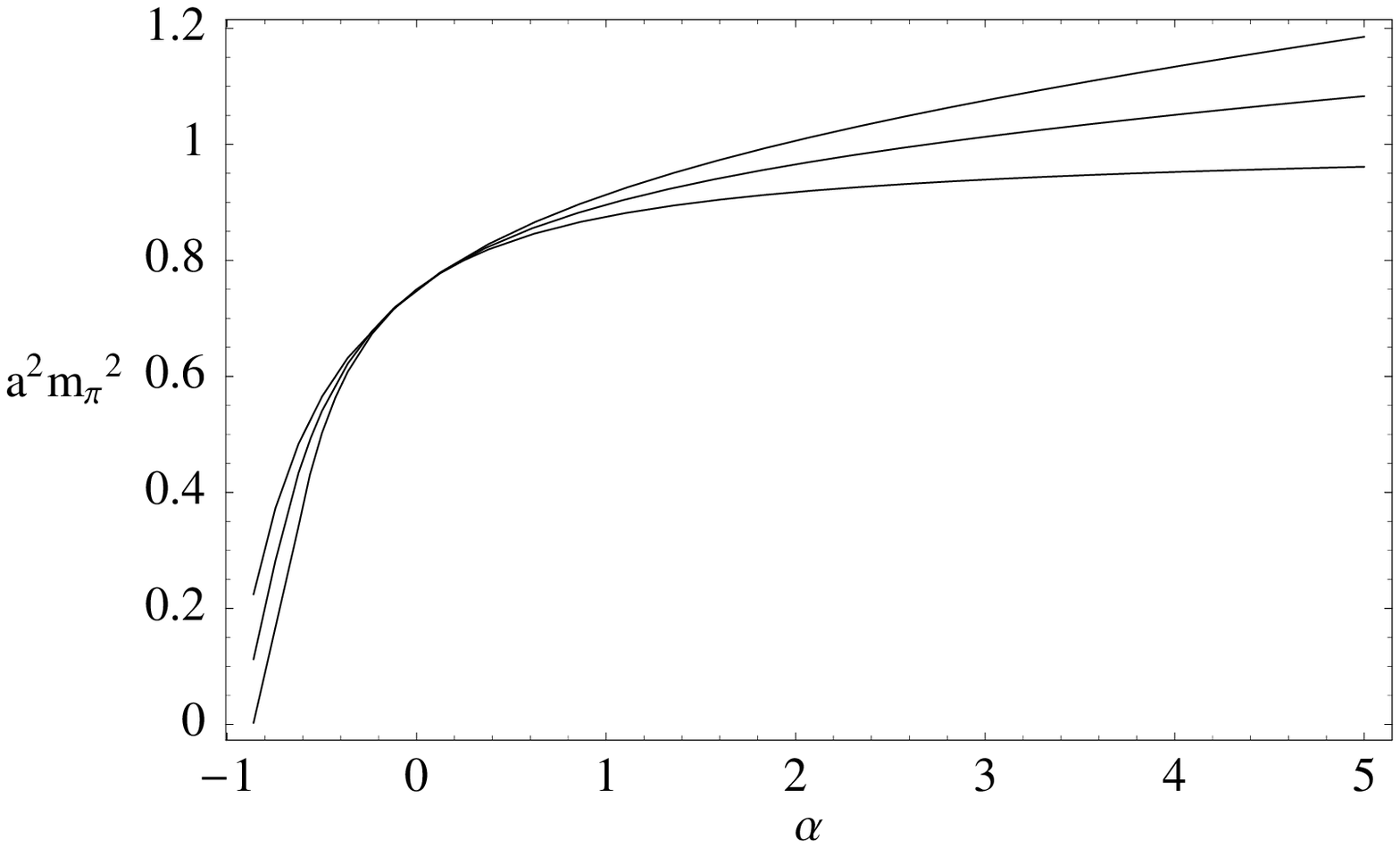,width=9cm}

\end{center}
\noindent {\bf Figure 1:} $a^2 m_\pi^2$ as a function of the
squashing $\a$ with, from bottom to top on the left of the graph,
$a \lambda = 0.1$, $10$ and $1000$.
\end{figure}

An interesting feature of figure 1 is that the value of $a^2
m_\pi^2$ on the round sphere, $\a=0$, is precisely the conformal
value for $\a=0$, $a^2 m_\pi^2 = 3/4$, for all values of the coupling. This
can be checked analytically using the expressions for the zeta
functions in Appendix A. Thus the theory with tree level conformal
mass term $a^2m^2=3/4$ posesses a self-consistent solution to the gap
equation, for any non-zero coupling $a\lambda$, precisely at
$a^2m_\pi^2=3/4$, which is again the conformal value! This can be understood
as follows. On ${\mathbb R}^3$ the critical $O(N)$ model at large $N$ is
obtained when $m_\pi^2$ vanishes. The round $S^3$ is conformally
equivalent to ${\mathbb R}^3$. Thus it is natural that $m_\pi^2$
should take the conformal value on the round sphere. In fact this
provides a check of our
analytically continued expressions and will have a nice dual
gravitational interpretation below.

We now turn to the strong and weak coupling results for the
theory which may also be interpreted as large and small volume
regimes.

\subsubsection{Weak coupling}

In the weak coupling regime we may solve for $m_\pi^2$
analytically. From the fact
that $\zeta(1) \sim 1/m_\pi^2$ as $m_\pi^2 \to 0$ and from the gap equation
(\ref{eq:massivesol}) it
follows that
\bea\label{eq:weakcoupling}
m_{\pi}^2 & \approx & \frac{- \lambda}{m^2 \text{Vol}{\mathcal{M}}} \quad
\text{for}\quad a \lambda \ll 1 \quad \text{if} \quad \a <
\frac{-3}{4} \nonumber \,, \\
m^2_{\pi} & \approx & m^2 + {\mathcal{O}}(\lambda) \,, \quad \text{for}
\quad a\lambda \ll 1 \quad \text{if} \quad \a > \frac{-3}{4} \,.
\eea
We can see that as $a\lambda \to 0^+$ the derivative of $m_\pi^2$
becomes discontinuous at $\a = -3/4$. This is illustrated in figure 2.
\begin{figure}[h]
\begin{center}
\epsfig{file=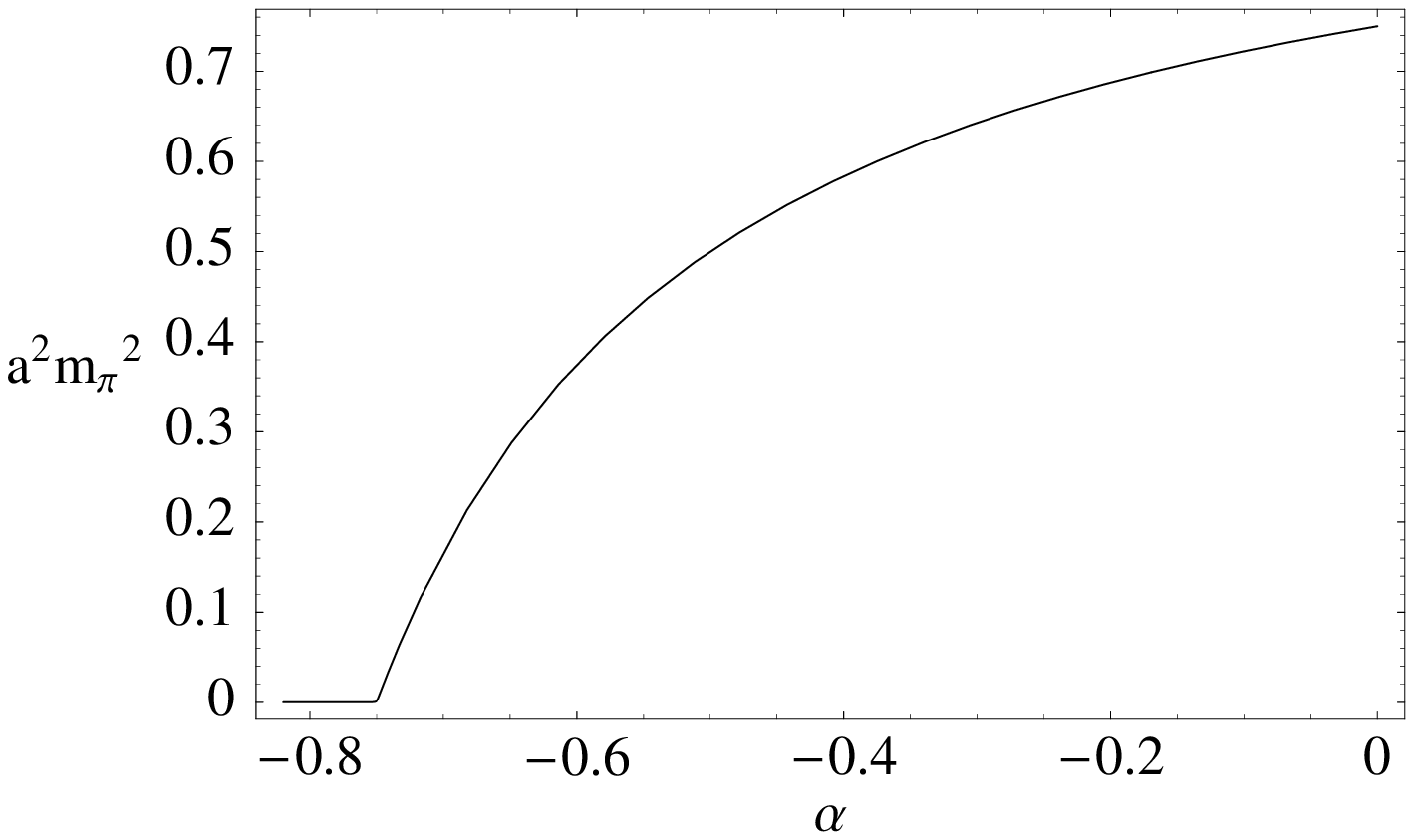,width=9cm}

\end{center}
\noindent {\bf Figure 2:} $a^2 m_\pi^2$ as a function of the
squashing $\a < 0$ and with $a\lambda \to 0^+$.
\end{figure}

The discontinuity suggests a phase transition at $\a = -3/4$ in the
extreme weak coupling limit $a\lambda \to 0^+$. Indeed $m_\pi^2 \to 0$
in this limit. The vanishing of the ``pion'' mass, $m_\pi^2$, is
reminiscent of symmetry breaking leading to massless Goldstone
bosons. However, on a compact space there will be no symmetry
breaking. Nevertheless, in the strict weak coupling limit there is a
new phase that appears for $\a < -3/4$. We know from (\ref{eq:sp0})
that at finite volume this must require $\phi^2 \to \infty$. Indeed
we will see below that this
phase transition is characterized by an order parameter, namely the 
condensate
$\lambda <\vec{\Phi}{\cdot}\vec{\Phi}>$ which vanishes for $\a<-3/4$
and is nonvanishing and of order $N$ for $\a>-3/4$.
However, in most of this paper we will be interested in the case of
finite coupling in which this phase transition is completely smoothed out.
The nonperturbative smoothing of phase transitions may be a more
general phenomenon of relevance in attempts to match perturbative phase
transitions with strong coupling results
\cite{Aharony:2005bq,Aharony:2003sx}.

\subsubsection{Strong coupling}

For any finite value of the coupling the gap
equation (\ref{eq:pionmass})
seems difficult to solve analytically. In the limit of infinite coupling, 
the gap
equation simply becomes
\be\label{eq:strongcoupling}
\zeta_{m_\pi^2}(1) = 0 \quad \text{as} \quad a\lambda \to \infty \,.
\ee
Again, this equation does not appear easy to solve for general $\alpha$ but
the roots can be determined numerically. However, it turns
out to be analytically tractable at large $\a$ leading to a rather
suggestive solution. Using the large $\a$
expression for the zeta function discussed in Appendix C we find that
$m_\pi^2$ grows linearly with $\alpha$
\be\label{eq:goldenmean}
\fbox{ $\displaystyle
a^2 m_\pi^2 \approx \frac{\a}{\pi^2} \ln^2
\left[\frac{1+\sqrt{5}}{2}\right] \quad \text{as} \quad \a \to \infty
\,. $}
\ee
The appearance of the golden mean in this expression is somewhat curious.
Interpreting $\sqrt\alpha$ as a temperature $T$, this equation gives
us the high temperature behaviour of the solution to the gap equation,
$m_\pi^2\sim T^2$. In fact, following the technique outlined in
Appendix C one obtains exactly the same expression as above for
the $O(N)$ model on $S^1 \times S^2$ in the limit of high temperature,
or shrinking $S^1$. In this $S^1 \times \R^2$ limit, the expression
(\ref{eq:goldenmean}) for the mass in terms of the golden mean was
found previously in the $O(N)$ model in \cite{Sachdev:1993pr}.

\subsection{Physics of the symmetric vacuum II: the free energy}

Using the solution to the large $N$ gap equation we compute
the value of the action of the model as a function of
$\alpha$. We may do this keeping the volume of the
squashed $S^3$ fixed while varying the squashing parameter. This has
the effect of removing the logarithmic dependence of the action
(\ref{eq:F}) on
the volume of the squashed sphere.
This action has a natural thermodynamical
interpretation as the free energy times the inverse temperature.
In figure 3 we plot the action of the symmetric vacuum (\ref{eq:F})
against $\a$ for strong coupling $a\lambda\rightarrow \infty$. The
qualitative form of the plot remains unchanged at finite values of the
coupling, although it does change for weak coupling.
\begin{figure}[h]
\begin{center}
\epsfig{file=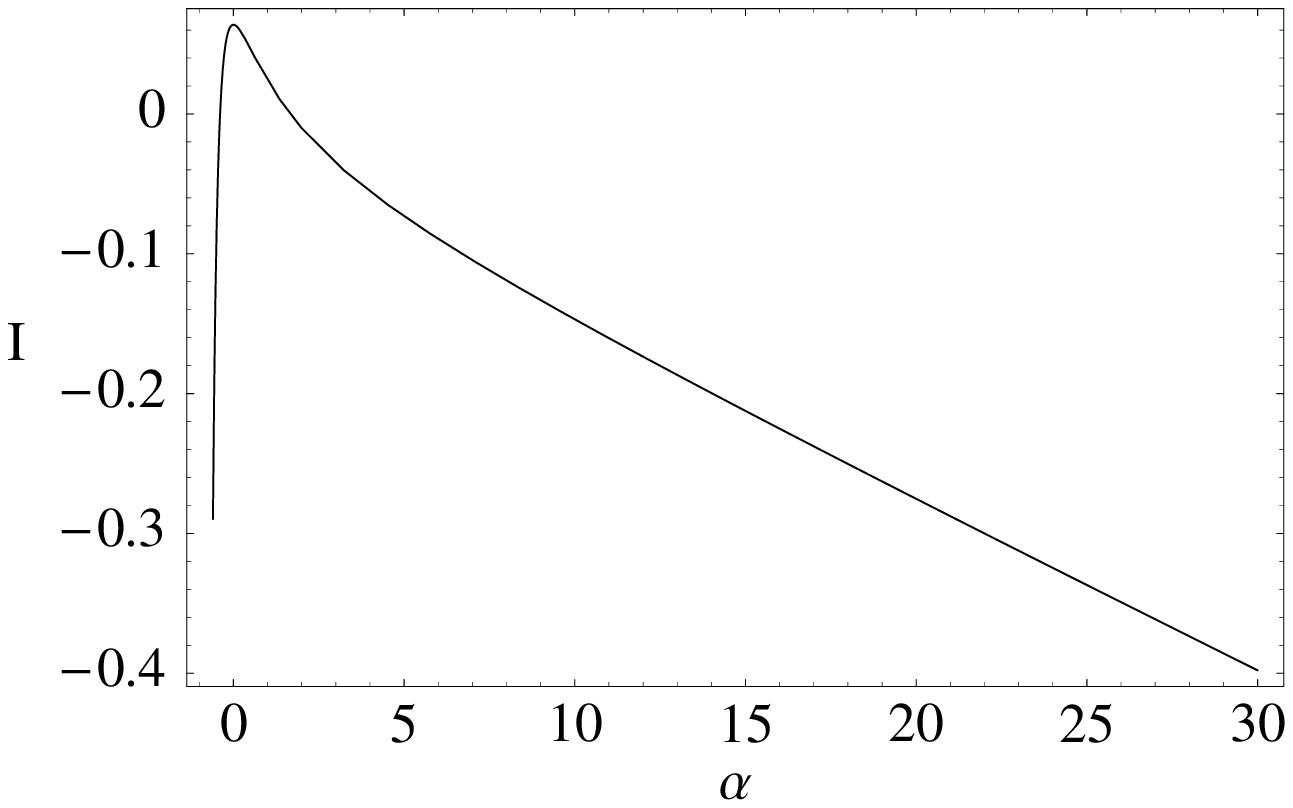,width=9cm}

\end{center}
\noindent {\bf Figure 3:} The field theory action $I$ as a function of the
squashing $\a$ with $a \lambda \rightarrow \infty$.
\end{figure}

The interesting features of this plot include the behaviour at
large positive $\a$, the behaviour near $\a = -1$ and the
presence of a maximum at $\a=0$.
Based on numerical analysis, we are able to deduce that near
$\alpha = -1$, the action goes like $I\sim -1/(1+\alpha)^2$. For large
positive $\alpha$ the action scales linearly with $\alpha$, which may
be shown analytically. We will see below that all three features are also
encountered in the dual gravitational theory.

We may find a closed expression for the free energy at large
squashing $\a$ using (\ref{eq:goldenmean}) for strong coupling and
(\ref{eq:weakcoupling}) for weak coupling. The techniques discussed
in Appendix C yield
\be\label{eq:f1}
\fbox{ $\displaystyle
\left. I\, \right|_{a\lambda \ll 1} \approx - \frac{N \zeta_R(3)}{8
  \pi^2}\, \a \quad \text{as} \quad \a \to \infty \,, $}
\ee
and also remarkably
\be\label{eq:f2}
\fbox{ $\displaystyle
\left. I\, \right|_{a\lambda \to \infty} \approx \frac{4}{5}
\left. I\, \right|_{a\lambda \ll 1} \quad \text{as} \quad \a \to \infty \,.
$}
\ee
In these expressions $\zeta_R(s)$ denotes the Riemann zeta
function. If one considers the squashed $S^1$ direction as a
nontrivially fibred temperature, then the linear dependence on $\a$ is
interpreted as a $T^2$ scaling with temperature.
Writing $F=IT$, this gives the usual
high temperature scaling of the free energy in a three dimensional
spacetime.

The factor of $4/5$ difference between the strong and weak coupling
regimes is reminiscent of the well known $3/4$ factor that
distinguishes the strong and weak coupling limits of the free energy of
${\mathcal{N}}=4$ super Yang Mills theory in four dimensions
\cite{Gubser:1996de}. In the present context we
are able to calculate both limits nonperturbatively within field
theory. The presence of $\zeta_R(3)$ in the free energies
(\ref{eq:f1}) and (\ref{eq:f2}) is generic for conformal field
theories on $S^1 \times \R^2$, the high temperature limit, see
\cite{Sachdev:1993pr} and references therein. That paper also finds a
factor of 4/5 in the free energy of the $O(N)$ model on $S^1 \times
\R^2$.

\section{The gravitational dual}
\label{sec:dual}

According to the original correspondence proposed by Klebanov and
Polyakov \cite{Klebanov:2002ja} the gravitational theory dual to the
three dimensional critical
$O(N)$ vector model is a higher spin gauge theory in
${\rm AdS}_4$ spacetime \cite{Fronsdal:1978rb, Fradkin:1987ks,
  Fradkin:1986qy}. The details of this bulk theory, and hence the
correspondence itself, have not yet been completely understood for
${\rm AdS}_4$ spacetime whose boundary in global coordinates is the
round three-sphere \cite{Petkou:2003zz}.
For the $O(N)$ model on squashed spheres, the dual gravitational
description should be a
higher spin gauge theory on bulk geometries whose conformal boundaries
are squashed three-spheres. We leave an in-depth study of the higher
spin theory to future work. Instead, we will test the exact
results from our field theory analysis against known semiclassical
results regarding the associated bulk geometries, without taking into
account the effect of any higher spin degrees of freedom.
We find a remarkably detailed qualitative agreement between the two 
pictures.

In general there are two four dimensional Riemannian geometries
with negative cosmological constant that have the squashed
three-sphere as conformal boundary. These are the AdS Taub-NUT and the AdS
Taub-Bolt spacetimes. In Einstein semiclassical quantum gravity there 
is a
first order phase transition between the two geometries at a critical
value of the squashing parameter
\cite{Chamblin:1998pz,Hawking:1998ct}. This phase transition is simply
the NUT charged version of the Hawking-Page transition for black
holes in AdS \cite{Hawking:1982dh}.
More concretely, one finds that
the AdS Taub-Bolt geometry only exists for $\a \geq
5+3\sqrt{3}\approx 10.2$  and the quantum gravity phase transition
itself occurs at $\a_\text{crit} = 6+2\sqrt{10}\approx 12.3$
\cite{Chamblin:1998pz,Hawking:1998ct}. The authors of
\cite{Chamblin:1998pz,Hawking:1998ct} computed the difference of
the actions associated to the AdS Taub-NUT and AdS
Taub-Bolt spacetimes and inferred the existence of a transition at the
point where the difference vanishes. More interestingly from our
point of view, in \cite{Emparan:1999pm,Mann:1999pc} the individual actions for
the AdS Taub-NUT and AdS Taub-Bolt geometries were obtained using a
boundary countertem technique inspired by the prescription
of \cite{Balasubramanian:1999re}. We quote the result of
\cite{Emparan:1999pm,Mann:1999pc} for the bulk action as a function of the
boundary squashing parameter
\be
I_{\rm TN}=-{6\pi\over GR } {(2\a+1)\over(\a+1)^2} \,,
\ee
where $G$ is Newton's constant and $R$ is the (negative) Ricci scalar for 
AdS
Taub-NUT. Note that we use $R$ to denote both the boundary and the
bulk scalar curvatures. This solution exists for the entire range of $\a$ 
and
dominates the partition function for $\a < 6+2\sqrt{10}$. In
particular near $\alpha =-1$ the bulk action behaves as $I_{\rm
  TN}\sim -1/(1+\a)^2$ which is precisely what we found in the strongly
coupled field theory. In addition $I_{\rm TN}$ has a maximum at
$\a=0$, mirroring our field theory result of figure 3.

For $\alpha>6+2\sqrt{10}$ the bulk action is that of AdS Taub-Bolt
\be
I_{\rm TB}= -{24\pi\over R G} \;(1+\a)^{-1/2}\;\left(m_b+{3\over
  4}\;r\;(1+\a)^{-1}-r^3\right) \,,
\ee
with
\be
m_b={1\over 2} r +{1\over 8 r} (1+\a)^{-1} +{1\over 2}\left(r^3-{3\over
  2}r(1+\a)^{-1} -{3\over 16 r}(1+\a)^{-2}\right) \,,
\ee
and
\be
r={1\over 6} (1+\a)^{1/2}\left(1+ 
\sqrt{1-12(1+\a)^{-1}+9(1+\a)^{-2}}\right)\,.
\ee

For large $\alpha$, the Bolt action is negative and grows linearly
\be\label{eq:ITB}
I_{\rm TB} \approx \frac{4 \pi}{9 G R} \alpha \quad \text{as} \quad \a \to 
\infty \,.
\ee
This linear
behaviour of the action translates into a $T^2$ scaling law with
temperature, and hence a $T^3$ scaling of the free
energy with temperature. This is essential in order to have an
interpretation in terms of a three dimensional boundary field
theory. Indeed, our field
theory result (\ref{eq:f2}) exhibits a negatively sloped linear dependence
on $\alpha$ for large squashings.
On the other hand,
the AdS Taub-NUT solution vanishes for large $\alpha$ as
$I_{\rm TN}\sim 1/\a$.

\begin{figure}[h]
\begin{center}
\epsfig{file=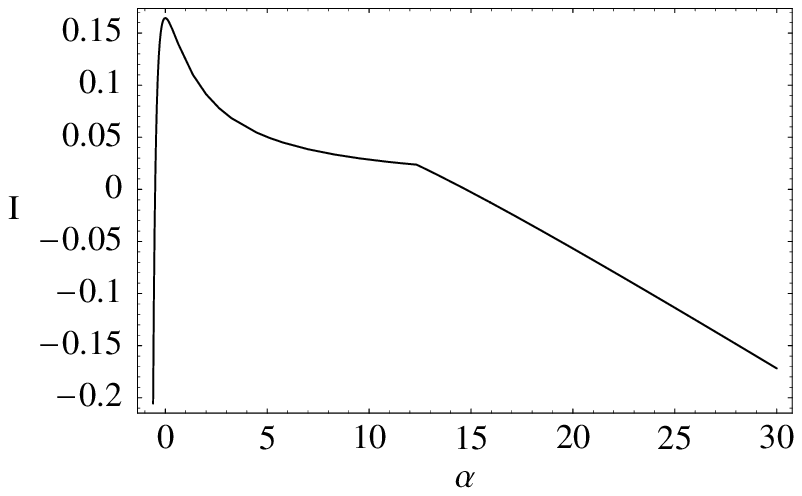,width=9cm}

\end{center}
\noindent {\bf Figure 4:} The bulk action $I$ as a function of the
squashing $\a$.
\end{figure}

In figure 4 we have plotted the bulk action as a function of
$\alpha$. We have normalised the action so that its slope at
large $\a$ agrees with the field theory result (\ref{eq:f2}).
There is an additive constant in the action which could be adjusted
to make the peak values of the actions match.
Comparing with figure 3, which was obtained from the strongly coupled field
theory, we clearly see the qualitative similarities at small and large
values of $\a$. However, it is clear that there is no field theory
analog of the Hawking-Page transition seen in the bulk. As we have pointed
out earlier, we can presumably attribute this to the fact that the
infinite massless higher
spin degrees of freedom have not been taken into account in the bulk
analysis.

We should emphasise that the qualitative behaviour of the field theory
at very weak coupling is quite different to the strong coupling
regime, particularly
for the negative values of $\a$ where the boundary curvature becomes
negative. At weak coupling we find that the action drops
rapidly for $\a<-3/4$ and scales as $-1/(1+\a)^{5/2}$ which is faster
than the gravity result which scales like $-1/(1+\a)^2$ near
$\a=-1$. This is fairly
straightforward to see following our treatment of the weak coupling
regime.
The strongly coupled field theory on the other hand, based on
numerical evidence, appears to match
the behaviour of the bulk theory near $\a=-1$. This suggests, as is usual in
AdS/CFT dualities, that the strongly coupled field theory is dual to
gravity on a weakly curved space. Below we will present another piece of
evidence in favour of this identification which pertains to the
behaviour of the field
theory condensate $\langle\vec\Phi\cdot\vec\Phi\rangle$ in the
strongly coupled field theory and its implications for the gravity dual.

Equating the bulk and boundary values for the effective action at large 
$\a$,
(\ref{eq:f2}) and (\ref{eq:ITB}), suggests the tentative dictionary
\be\label{eq:dict}
-R G = \frac{\pi^3 40}{9 \zeta_R(3)} \frac{1}{N} \,.
\ee
This formula passes the immediate test of giving weak curvatures when
$N$ is large. It might be compared with the well known result $R G
\sim 1/N^{3/2}$ for the $AdS_4\times S^7$ version of the
correspondence. The appearance of $\zeta_R(3)$ in (\ref{eq:dict}) is
curious and possibly tantalising given that $\zeta_R(3)$ also appears
in a computation of $\a'$ corrections to the IIB effective supergravity 
action and hence the bulk free energy \cite{Gubser:1998nz} for the
$AdS_5 \times S^5$ version of the correspondence. In general
we expect the coefficient in (\ref{eq:dict}) to depend on the
background. The same $R G \sim 1/N$ relation was found for the
Klebanov-Polyakov correspondence on $AdS_4$ \cite{Petkou:2003zz}.

\subsection{Negative curvature of the boundary}

Let us consider the physics associated with the sign change in
the scalar curvature of the boundary at $\a = -3/4$. From our comments
above,
the dual geometry to the squashed three-sphere in this regime is
the AdS Taub-NUT spacetime which may be written as
\be\label{eq:taub}
ds^2 = \frac{1}{k^2(1-r^2)^2} \left[\frac{4(1+\a r^2)}{1+\a r^4} dr^2
  + r^2 (1+\a r^2) (\sigma_1^2 + \sigma_2^2) + r^2 \frac{1+\a
    r^4}{1+\a r^2} \sigma_3^2  \right] \,,
\ee
where the cosmological constant is $\Lambda = -3k^2$ and the range of
the radial coordinate is $0 \leq r < 1$. We have already argued above
that the weakly curved gravitational description appears to
be dual to the strongly coupled $O(N)$ model. Therefore
we expect that the fact that a
stable symmetric phase continues to exist for $\a < -3/4$ at strong
coupling should translate into the bulk spacetime remaining stable in
this regime. We have also seen that there is a
phase transition at $\a = -3/4$ for arbitrarily weak coupling. However, we
expect this to be a strong curvature effect in the gravitational
dual and thus beyond a classical gravitational computation.

In string theory duals to field theories there are generically two
types of instabilities that can arise when the conformal boundary has
negative scalar curvature. For a summary of these, see
\cite{Kleban:2004bv}.
The first is a nonperturbative instability due to the nucleation of
BPS branes. The second is a perturbative instability due to negative
modes of fluctuations about the solution. It is unclear whether the
bulk higher spin gauge theory will contain objects analogous to BPS
branes, so we will focus on the perturbative possibility.

The perturbative stability of the AdS Taub-NUT spacetime against
fluctuations of a scalar field was considered in
\cite{Kleban:2004bv}. That paper was primarily interested in scalar fields
with a mass saturating the Breitenlohner-Freedman bound, as these arise from
the
compactification of eleven dimensional supergravity to four
dimensions. The higher spin gauge theory that is conjectured to be dual to
the $O(N)$ model has instead a conformally coupled scalar field, $\varphi$.
It is simple to adapt the analysis of \cite{Kleban:2004bv} to this case. We
are
interested in whether the bulk equation
\be
- \Box \varphi + \frac{R}{6}\varphi = \beta \varphi \,,
\ee
has solutions with negative $\beta$. This equation may be separated
and converted into Schr\"odinger form
\be
- \frac{d^2 \chi}{dr_*^2} + V(r_*) \chi = \beta \chi \,,
\ee
using the transformations
\be
\chi = (g g^{rr})^{1/4} \varphi \,, \quad dr = (g^{rr})^{1/2} dr_* \,,
\ee
where $g$ is the determinant of the metric (\ref{eq:taub}). It is then
easy to see that the potential $V(r_*)$, is everywhere positive for
all values of
the squashing $\alpha$. Therefore $\beta$ cannot be negative and the
spacetime is perturbatively stable.

One should also check stability against the other fields in the
theory, such as metric fluctuations. However, stability of the bulk
would be consistent with our field theory finding that the symmetric
phase remains stable at finite coupling as we go to the regime of
parameter space with a negatively curved boundary.

\subsection{The condensate $\langle\vec{\Phi}{\cdot}\vec{\Phi}\rangle$}

Finally, we turn to an analysis of the condensate
$\langle\vec{\Phi}{\cdot}\vec{\Phi}\rangle$ in field theory and its
implications for bulk physics. We will also argue that the condensate can
be used to construct an order parameter for the extreme weak coupling phase
transition discussed in section 3.1.1.

From the defining equations for the composite operator $\sigma$ and
pion mass $m_\pi^2$, (\ref{eq:aux2}) and (\ref{eq:mass}), we have that
\be\label{eq:condensate}
\frac{1}{N}\langle\vec{\Phi}{\cdot}\vec{\Phi}\rangle = \frac{1}{\lambda}
\left(m^2_\pi - m^2
\right) \,.
\ee
The Klebanov-Polyakov duality \cite{Klebanov:2002ja} then relates the
vacuum expectation value
$\langle\vec{\Phi}{\cdot}\vec{\Phi}\rangle$ to the normalisable mode of
the conformally coupled bulk scalar field $\varphi$.

Let us firstly consider the strong coupling limit.
We saw in equation (\ref{eq:strongcoupling})
that $m^2_\pi - m^2 = {\mathcal{O}}(1)$ as $a\lambda \to \infty$, that
is, $m_\pi^2$ remains finite as we take the strong coupling limit. Thus
we have
\be
\frac{1}{N}\langle\vec{\Phi}{\cdot}\vec{\Phi}\rangle \;\sim\;
{\mathcal{O}}\left(\frac{1}{\lambda}\right) \quad \text{as} \quad
a\lambda \to \infty \,.
\ee
This implies that at strong coupling, $a \lambda \to \infty$, the
condensate vanishes. If we identify the strong coupling regime of the
$O(N)$ model with the weakly curved gravitational dual, then this
result translates into the (true) statement that the bulk field $\varphi$ is
not turned on in the AdS Taub-NUT and AdS Taub-Bolt geometries.

The nonvanishing of the condensate at finite coupling
then translates into the prediction that the bulk must allow for a
nonvanishing value for $\varphi$ in these cases. It is likely that these 
will just be
new backgrounds with a nonvanishing $\varphi$ profile turned on
self-consistently due to higher curvature corrections. Figure 5 shows a plot 
of
$\langle\vec{\Phi}{\cdot}\vec{\Phi}\rangle$ as a function of the
squashing at two
values of the coupling.
\begin{figure}[h]
\begin{center}
\epsfig{file=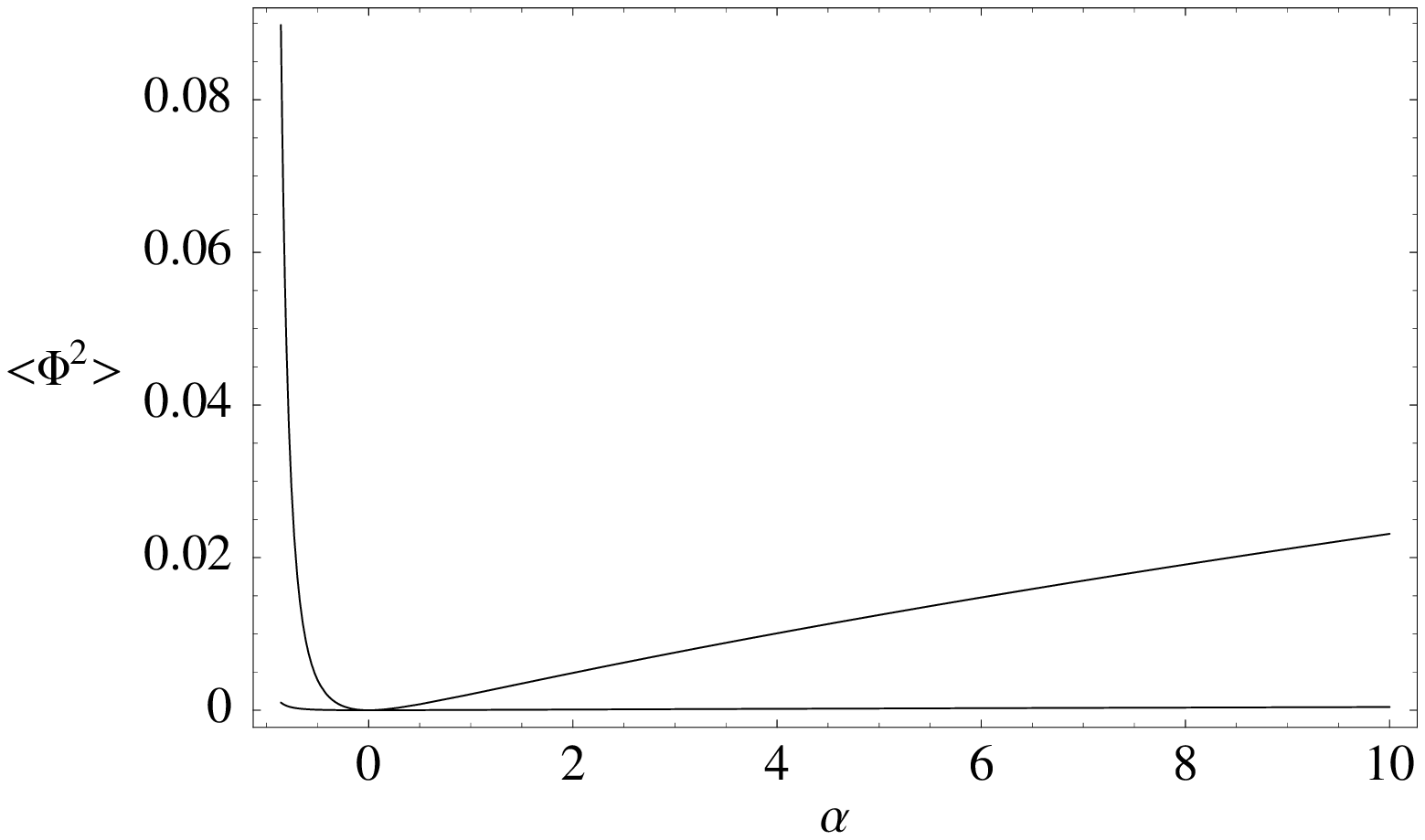,width=9cm}

\end{center}
\noindent {\bf Figure 5:} The condensate
$\langle\vec{\Phi}{\cdot}\vec{\Phi}\rangle/N$ as a function of the
squashing $\a$ with $a \lambda = 10$ (top) and $a \lambda = 1000$
(bottom).
\end{figure}

One remarkable aspect of figure 5, which was anticipated already in
figure 1, is that the condensate vanishes for the round sphere $\a =
0$ for all values of the coupling. The dual implication of this fact
is that when the bulk geometry is simply AdS, as opposed to AdS
Taub-NUT or AdS Taub-Bolt, then it does not receive
corrections at finite curvatures which turn on the scalar
$\varphi$. Such protection against corrections seems
plausible, given the maximal symmetry of AdS,
and mirrors known results for other realisations of the
AdS/CFT correspondence, see for instance \cite{Banks:1998nr} for the
$AdS_5\times S^5$ version of the correspondence.

At weak coupling the behaviour of the condensate (\ref{eq:condensate})
follows immediately from our previous results (\ref{eq:weakcoupling}).
In particular these imply that we may use the condensate as an order
parameter for the symmetry breaking phase transition occuring at $\a=-3/4$
as $a \lambda
\to 0^+$
\be
\lim_{a\lambda \to 0^+} \left[ \frac{\lambda}{N}
  \langle\vec{\Phi}{\cdot}\vec{\Phi}\rangle
  \right] = \left\{
\begin{array}{l}
{\mathcal{O}}(1) \quad \text{for} \quad \a < -3/4 \,, \\
0 \quad \text{for} \quad \a > -3/4 \,.
\end{array}
\right.
\ee

\section{Some open questions}

Motivated by the duality conjectured by Klebanov and Polyakov
\cite{Klebanov:2002ja},
in this paper we have investigated and solved various aspects of the large
$N$ limit of the $O(N)$ vector model on a squashed three-sphere.
Perhaps surprisingly we have found nontrivial agreement
between our field theory results and the semiclassical gravitational
physics of the proposed dual geometries: AdS Taub-NUT and AdS Taub-Bolt.

There are various open ends to this work. The most obvious open
question is the fate of the Hawking-Page transition in the bulk theory
when massless higher spin degrees of freedom are taken into account. The field
theory results suggest a smoothing out of the transition.

Within field theory, one limit which we have not considered in detail
is the $\a \to -1$ limit. This is a rather singular limit, in which one dimension
of the squashed sphere
decompactifies. Nonetheless, perhaps there is interesting physics to
be discovered. Our results near this limit have been mostly
numerical. We have observed numerically that the strongly
coupled field
theory action in this regime scales in the same way as the action for the
AdS Taub-NUT solution, with $I\sim -1/(1+\alpha)^2$. An
analytical solution of the field theory problem in this regime may
shed further light upon the agreement between field theory and gravity.

Finally, it would be interesting to know if there are deeper reasons
underlying certain amusing numerical coincidences we observed in the exact
solution to the strongly coupled theory. 
Namely the appearance of the
golden mean in the solution (\ref{eq:goldenmean}) to the gap equation at high
temperature, or large
squashing, and the appearance of 4/5 as the ratio of
the weak and strong coupling limits of the free energy (\ref{eq:f2})
at large squashing. These results are essentially
due to the $O(N)$ model on $S^1 \times \R^2$ \cite{Sachdev:1993pr}.
\vspace{0.5cm}

{\bf Acknowledgements}: We would like to thank Nick Dorey, Gary
Gibbons, Asad Naqvi and Dave Tong for several enjoyable discussions
during the course of this work. We would also like to thank Anastasios
Petkou for drawing our attention to some of the relevant older
literature. SAH is supported by a research
fellowship from Clare College, Cambridge. SPK acknowledges support
from a PPARC Rolling Grant.

\startappendix

\Appendix{Analytic continuation of the zeta function}

The zeta function we wish to analytically continue is
\be
\frac{1}{\mu^{2s}}\zeta(s) = \sum_{l=1}^{\infty} \sum_{q=0}^{l-1} \frac{l
a^{2s}}{\left[l^2+\a
  (l-1-2q)^2 + a^2m^2_\pi - 1\right]^s} \equiv \sum_{l=1}^{\infty}
\sum_{q=0}^{l-1} f(q,l)\,.
\ee
The method used has two steps, the first turns the $q$ sum into an
integral and the second deals with the $l$ summation
\cite{Shen:1986jr}.

The Abel-Plana formula converts the $q$ sum into integrals.
To apply this formula we need to know the location of the branch
points of $f(q,l)$. These are
\be
q_\pm = \frac{l-1}{2} \pm \frac{i}{2} \left(\frac{l^2-1+a^2
  m^2_\pi}{\a} \right)^{1/2} \,.
\ee
We see that the sign of $\a$ is important for determining the position
of the branch points.
Let us firstly consider the case $\a < 0$. In this case, the branch
points are on the real axis and the Abel-Plana formula is
\be\label{eq:abel}
\sum_{q=0}^{l-1} f(q,l) = \int_{-\frac{1}{2}}^{l-\frac{1}{2}} f(t,l) dt - 2i
\int_0^{\infty} \frac{dt}{e^{2\pi t}+1} \left[f(1/2+it,l)-f(1/2-it,l)
\right]\,,
\ee
so long as the branch points $q_\pm$ are outside the range of
integration $(-1/2,l-1/2)$. For a given value of $\a$ this
may be achieved by curtailing the range of $l$. We will restrict
attention to $-8/9 < \a$ which will allow us to use the
previous expression (\ref{eq:abel}) for all values $l \geq 3$. The
remaining terms in the sum corresponding to $l=1,2$ are then added
to the final expression to obtain
\be\label{eq:appendzeta}
\frac{1}{\mu^{2s}}\zeta(s) = \frac{1}{m^{2s}_\pi} + \frac{4a^{2s}}{(3+a^2
m^2_\pi+\a)^s}
+ a^{2s} \int_0^1 \frac{F(y) dy}{(1+\a y^2)^s} +
\frac{a^{2s}}{(1+\a)^s} \int_0^{\infty} \frac{G(y) dy}{e^{2\pi y} + 1} \,,
\ee
with
\bea\label{eq:genFG}
F(y) & = & \sum_{l=3}^{\infty} \frac{l^2}{(l^2-A^2)^s} \,, \\
G(y) & = & 2i \sum_{l=3}^{\infty} \left[\frac{l}{([l+iB]^2-E^2)^s} -
\frac{l}{([l-iB]^2-E^2)^s} \right] \,,
\eea
where $A,B,E$ are given by
\bea\label{eq:ABE}
A^2 & = & \frac{1-a^2 m^2_\pi}{1+\a y^2} \,, \nonumber \\
B & = & \frac{2\a y}{1+\a} \,, \nonumber \\
E^2 & = & \frac{1-a^2 m^2_\pi}{1+\a} + \frac{4\a y^2}{(1+\a)^2} \,.
\eea
Note that $E^2$ changes sign at an intermediate value of $y$.

We now continue the sums in (\ref{eq:genFG}) using the
Sommerfeld-Watson transformation. For the case of $F(y)$ we have
\be
F(y) = \frac{i}{2} \int_{C_1} dz \frac{z^2 \cot\pi z}{(z^2-A^2)^s}
\,,
\ee
where the contour $C_1$ is shown in figure 6. By considering
sufficiently large $s$ we may rotate this contour into $C_2$, also
shown in figure 6, to obtain
\be\label{eq:finalF}
F(y) = \frac{i}{2} \int_{C_2} dz \frac{z^2 \cot\pi z}{(z^2-A^2)^s}
\,.
\ee
\begin{figure}[h]
\begin{center}
\epsfig{file=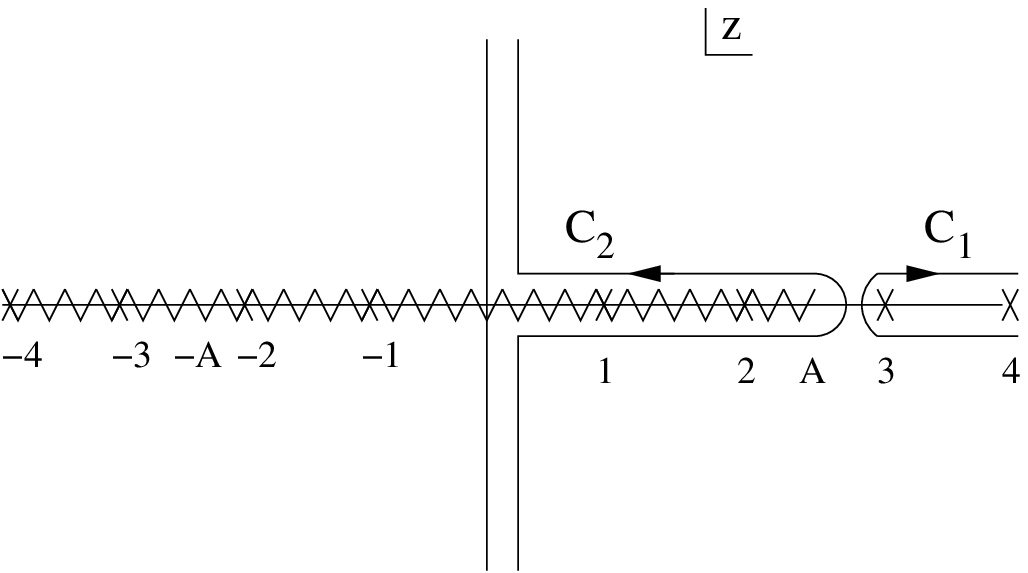,width=9cm}

\noindent {\bf Figure 6:} Change of contours for the analytic
continuation.
\end{center}
\end{figure}

This contour integral will require one more analytic continuation
along the imaginary axis using
\be
\coth\pi t = 1 + \frac{2}{e^{2\pi t}-1} \,.
\ee
We are interested in $F(y)$ at $s=0,1$ and in its
derivative with respect to $s$ at $s=0$. One finds, after paying
due attention to the branch cuts, the values
\bea
\left. F(y) \right|_{s=1} & = & \frac{-\pi A}{2} \cot \pi A + \sum_{p=1}^2
\frac{p^2}{A^2-p^2} \,, \\
\left. F(y) \right|_{s=0} & = & -5 \,.
\eea
The derivative at $s=0$ may be shown to be
\bea
\left. \frac{d}{ds} F(y) \right|_{s=0} & = & - \Re \left[ \frac{i \pi
  A^3}{3} + A^2 \ln \left(1-e^{-2i \pi A} \right) \right. \nonumber \\
& + & \left. \frac{iA}{\pi} \text{Li}_2\left(e^{-2i\pi A}\right)
+ \frac{1}{2 \pi^2} \text{Li}_3\left(e^{-2i\pi A}\right) \right]
+ \sum_{p=1}^2 p^2 \ln|A^2-p^2| \,.
\eea
In this expression $\text{Li}_2$ and $\text{Li}_3$ denote the
dilogarithm and trilogarithm functions, respectively.
It is sometimes simpler to avoid the polylogarithm functions. An
alternative expression for when $A$ is real is
\be
\left. \frac{d}{ds} F(y) \right|_{s=0} = - \pi P \int_{0}^{A} t^2
\cot \pi t dt - \frac{\zeta_R(3)}{2\pi^2} + \sum_{p=1}^2 p^2 \ln|A^2-p^2|
\,,
\ee
whilst if $A$ is imaginary we can write
\be
\left. \frac{d}{ds} F(y) \right|_{s=0} = \int_{0}^{|A|}
\frac{2\pi t^2}{e^{2\pi t} - 1} dt + \frac{i \pi A^3}{3}
- \frac{\zeta_R(3)}{2\pi^2} + \sum_{p=1}^2 p^2 \ln|A^2-p^2|\,.
\ee
In these expressions $\zeta_R(s)$ is the Riemann zeta function and $P$
denotes the Cauchy principal value of the integral.

Exactly the same manipulations are then performed for $G(y)$. The
contours $C_1$ and $C_2$ are qualitatively as before, except that
now the branch cuts emanate from the points $z=\pm E \pm i B $.
The values are now
\bea
\left. G(y) \right|_{s=1} & = & \frac{2\pi}{E} \frac{E\sinh (2\pi B) -
B\sin(2\pi
  E)}{\cosh(2\pi B) - \cos(2\pi E)} - \sum_{p=1}^2 \frac{8 B
p^2}{|E^2+B^2-p^2+i2pB|^2} \,, \\
\left. G(y) \right|_{s=0} & = & 0 \,.
\eea
The derivative is given by
\bea\label{eq:gd0}
\left. \frac{d}{ds} G(y) \right|_{s=0} & = &  \sum_{p=1}^2 i2p
\ln\frac{(p+iB)^2-E^2}{(p-iB)^2-E^2} - 4\pi \int_0^{\infty} \frac{B \cosh
2\pi B  -
  t\sinh 2\pi B -B e^{-2\pi t}}{\cosh 2\pi B - \cosh 2\pi t} dt \nonumber \\
& + & \sum_{\pm} \pm \left[ -2i E \ln \left(1- e^{2\pi[\pm
      B + iE]} \right) + \frac{1}{\pi} \text{Li}_2 \left(1 - e^{\pm 2\pi
    B} \right) \right. \nonumber \\
& - & \left. \frac{1}{\pi} \text{Li}_2 \left(1 - e^{2\pi[\pm
      B + iE]} \right) \right] + 2B \ln \frac{\cosh 2\pi
  B - 1}{\cosh 2\pi B - \cos 2\pi E} \,.
\eea
In this last expression, $\text{Li}_2(z)$ is the dilogarithm function.
In evaluating these logarithms and dilogarithms, one should be careful
to keep track of the phases of the complex numbers as $E$ increases.
In fact, the last two lines of (\ref{eq:gd0}) are given by
\be
- 4\pi \int_0^{E} \frac{B \sin 2\pi t - t \sinh 2\pi B}{\cosh 2\pi B -
\cos 2\pi t} dt \,,
\ee
when $E$ is real and by
\be
-4\pi |E| B + 4\pi \int_0^{|E|} \frac{B \cosh 2\pi B - t \sinh 2\pi B
  - B e^{-2\pi t}}{\cosh 2\pi B - \cosh 2\pi t} dt \,,
\ee
when $E$ is imaginary. These integral expressions are easier to
evaluate numerically than the dilogarithms.

When $\a > 0$ the branch points $q_\pm$ are not on the real
axis and $\Re q_\pm = (l-1)/2$ falls within the range of integration
in (\ref{eq:abel}). Application of the Abel-Plana formula is as for $\a < 0$
except
that there is an extra contribution to the zeta function
(\ref{eq:appendzeta}) given by an integral along the
branch cuts, which we take to run from $q_\pm$ to $\pm i \infty$,
respectively \cite{Shen:1986jr}. The extra term is
found to be
\be\label{eq:appendH}
a^{2s} H = - \frac{4 a^{2s}}{(4 \a)^s} \sin\pi s \sum_{l=3}^{\infty}
\int_P^{\infty} \frac{l}{[1+(-1)^l e^{2\pi y}]}
\frac{dy}{(y^2-P^2)^s}\,,
\ee
where
\be\label{eq:tempe}
P = \frac{1}{2} \left(\frac{l^2+ a^2 m^2_\pi - 1}{\a} \right)^{1/2} \,.
\ee
Evaluating this expression for $s=1$ one finds
\be\label{eq:defh}
\left. a^2 H \right|_{s=1} = - \sum_{l=3}^{\infty} \frac{\pi a^2 l}{2\a P}
\frac{1}{[1+(-1)^l e^{2\pi P}]} \,.
\ee
Whilst the derivative at $s=0$ is
\be\label{eq:temp}
\left. \frac{d}{ds} H \right|_{s=0} = - \sum_{l=3}^{\infty} 2 l \ln
\left[1 + (-1)^l e^{-2\pi P} \right] \,.
\ee
The sums in the last two expressions are convergent and may be
evaluated numerically. The last expression clearly displays the
connection to a theory at finite temperature for $\a>0$. If we
interpret $\sqrt \a$ as a temperature, the above expression reduces to
the finite temperature free energy  associated to a field theory on a
two sphere, along with certain parity reversal operations. 

\pagebreak

\Appendix{A check of the zeta function}

In \cite{Critchley:1982ch} a series expansion in $\a$ is given for the
zeta function on a squashed three-sphere for the value of the mass
$m_\pi^2 = R/6$ (note that this is not the conformal value in three
dimensions). The result is
\bea
\zeta(s) & = & (a\mu)^{2s} \left[ \zeta_R(2s-1) + \frac{\a}{3} s
\zeta_R(2s-2) \right. \\
& + & \left. 4\a^2 \left( \frac{1}{12} s \zeta_R(2s) + 2 s (s+1)
\left[\frac{1}{80} \zeta_R(2s-2) - \frac{1}{36}\zeta_R(2s) +
  \frac{1}{45} \zeta_R(2s+2) \right] \right)
\right] + \cdots \,. \nonumber
\eea

We have checked our expressions against this expansion and found very
good agreement at each order in $\alpha$. The agreement is rather
nontrivial as it involves very precise cancellations between the terms
involving $F$ and $G$ in the zeta function (\ref{eq:appendzeta}). It
does not however provide a check of the $H$ term (\ref{eq:appendH})
as this depends nonperturbatively on $\a$. However we have derived
this remaining term several times and furthermore it agrees precisely with
the
contribution calculated in \cite{Shen:1986jr}.

\Appendix{The zeta function at large squashing}

The limit $\a \to \infty$ of the zeta function may be treated
directly because one of the sums may be turned into an integral in this
limit. The zeta function sum (\ref{eq:sum}) may be rewritten as
\be
\zeta(s) = \frac{(a\mu)^{2s}}{\a^s} \sum_{k=-\infty}^{\infty}
\sum_{l\in |k|+1+2\N} \frac{l}{\left(\frac{l^2-1}{\a} + k^2 + \frac{a^2
    m_\pi^2}{\a} \right)^s} \,.
\ee
In the limit $\a \to \infty$, we may define $x=l/\a^{1/2}$ and let
\be
\sum_{l\in |k|+1+2\N} \; \to \; \frac{\a^{1/2}}{2}
\int_{(|k|+1)/\a^{1/2}}^{\infty} dx \,.
\ee
This treatment is only allowed for the $k=0$ terms if $a^2 m_\pi^2/\a$
remains finite as $\a \to \infty$. We will see that indeed this is the
case at strong coupling, $a\lambda \to \infty$.
Doing the $x$ integral, the zeta function becomes
\be\label{eq:largealpha}
\zeta(s) = \frac{(a\mu)^{2s}}{4 \a^{s-1}} \frac{1}{(s-1)}
\sum_{k=-\infty}^{\infty} \frac{1}{\left(k^2 + \frac{a^2
    m_\pi^2}{\a}\right)^{s-1}} \quad \text{as} \quad \a \to \infty \,.
\ee

To find the behaviour of $a^2 m_\pi^2$ at strong coupling and as $\a
\to \infty$ we need to solve (\ref{eq:strongcoupling}), that is
$\zeta(1)=0$. We may evaluate $\zeta(1)$ by analytic continuation of
(\ref{eq:largealpha}) using the techniques of appendix A. The result
is
\be
\zeta(1) =  - \frac{a^2}{2} \left[\frac{\pi a m_\pi}{\a^{1/2}} +
  \ln \left(1- e^{\displaystyle -2\pi a m_\pi/\a^{1/2}}  \right)
\right] \quad \text{as} \quad \a \to \infty \,.
\ee
Setting this expression to zero one obtains equation
(\ref{eq:goldenmean}) in text
\be\label{eq:themass}
a^2 m_\pi^2 = \frac{\a}{\pi^2} \ln^2
\left[\frac{1+\sqrt{5}}{2}\right] \quad \text{as} \quad \a \to \infty
\,.
\ee

To compute the free energy in this limit, we need to evaluate
$\zeta'(0)$. By the usual methods we obtain
\be
\zeta'(0) = \frac{\a}{2} \left[\frac{2 \pi (a m_\pi)^3}{3 \a^{3/2}} +
  \pi \int_{a m_\pi/\a}^{\infty} \frac{2 (t^2 - a^2
    m_\pi^2/\a)}{e^{2\pi t} - 1} \right] \quad \text{as} \quad \a \to \infty
\,.
\ee
The integral may be expressed in terms of dilogarithms and
trilogarithms. However, for the particular value of the mass given by
(\ref{eq:themass}) the expression simplifies dramatically to give the
strong coupling result
\be
\zeta'(0) = \frac{\zeta_R(3)}{5 \pi^2} \a \quad \text{as} \quad \a \to
\infty
\,,
\ee
which immediately leads to equation (\ref{eq:f2}) in the text. We have
checked numerically that these strong coupling results actually match
our analytically continued expressions involving $F, G$ and $H$ in
(\ref{eq:appendzeta}) in the large $\a$ regime. The matchings rely on
delicate cancellations between the $F$, $G$ and $H$ terms and thus
provide a nontrivial check of our analytic continuations.
The weak
coupling result
(\ref{eq:f1}) is obtained similarly, but in this case $a^2 m_\pi^2/\a
\to 0$ as $\a \to \infty$ and one has to treat the $k=0$ sum separately.

\end{document}